# Spatially Separated Self-Assembled Silver Dendrites for Highly Sensitive SERS Applications

D. V. Yakimchuk[1], E. Yu. Kaniukov[1], S. I. Lepeshov[2], V. D. Bundyukova[1], S. E. Demyanov[1], G. M. Arzumanyan[3,4], N. V. Doroshkevich[3], K. Z. Mamatkulov[3], A. Alù[5,6,7], A. Bochmann[8], M. Presselt[9], O. Stranik[9], A. Krasnok[2,5], and V. Sivakov[9*]

[1]*Scientific-Practical Materials Research Center NAS of Belarus, Minsk, Belarus*

[2]*ITMO University, St. Petersburg, Russian Federation*

[3]*Joint Institute for Nuclear Research, Dubna, Russian Federation*

[4]*Dubna State University, Dubna, Russian Federation*

[5]*Photonics Initiative, Advanced Science Research Center, United States*

[6]*Physics Program, Graduate Center City University of New York, United States*

[7]*Department of Electrical Engineering, City College of New York, United States*

[8]*Ernst-Abbe-Hochschule Jena, Jena, Germany*

[9]*Leibniz Institute of Photonic Technology, Jena, Germany*

E-mail: vladimir.sivakov@leibniz-ipht.de

## Abstract

Even after more than four decades of rapid-pace advances, the adoption of Surface-Enhanced Raman Scattering (SERS) remains limited, due to the difficulties in fabrication of effective and affordable nanostructured surfaces or templates. From this point-of-view, self-assembly techniques of SERS substrates formation allowing fabrication of highly branched dendritic silver nanostructures, which are very promising for biosensor applications. However, these techniques suffer from the lack of control over the resulting structures. Herein, we propose an electroless wet-chemical approach for highly reproducible fabrication of spatially separated plasmonic active silver dendrites in a porous matrix of Si/SiO$_2$ template via controlled self-assembly of silver in a limited volume. The proposed approach allows fabrication of spatially separated silver dendrites with sub-10-nm gaps between neighboring branches, which possessing strong electric field enhancement in a wide spectral range. The fabricated silver dendritic structures enable a SERS enhancement factor of ~$10^8$ and the analyte detection limit of ~$10^{-15}$ M, which is at the level of single molecule sensitivity. The comparison of simulation results with SERS experiments allows to distinguish the presence of electromagnetic and chemical contributions, which have a different effect at various analyte concentrations.

**Introduction**

Raman spectroscopy, consisting in measuring the inelastic light scattering from a molecule with quantized vibrational signature, is a highly innovative technique of modern biomedical analysis, clinical and biological studies[1]. These areas have been experiencing a fast development, which has been continuing for more than 40 years after the discovery that inherently weak Raman signals can be significantly enhanced for molecules adsorbed onto a rough metal surface. This effect, which can be attributed mainly to strong electric field enhancement around subwavelength metal roughness, is called Surface-Enhanced Raman Scattering (SERS)[2–5] and has been demonstrated for various plasmonic structures, including plasmonic nanoantennas[6], dimers with sub-10-nm gaps[7], disordered nanoparticle clusters, metamaterials, and metasurfaces[8].

Electromagnetic mechanisms usually provide the primary role in SERS amplification. Excitation of surface plasmons, i.e., coherent oscillations of free electrons at the metal-dielectric boundaries, causes regions of a strong electric field (hot spots) arising near metal nanostructures and at the junctions between them[9–11]. The SERS enhancement factor caused by this electric field localization is roughly proportional to the fourth power of the electric field amplitude. Subsequent analysis has shown that SERS is inherent not only to metals, such as copper (Cu)[12], silver (Ag)[13], and gold (Au)[14] but also to resonant all-dielectric nanostructures,[15] demonstrating the universal nature of this effect, associated with the local electric field enhancement. However, in some cases the chemical contribution is a factor responsible for the sharp growth of the Raman scattering cross-section due to the interaction of the analyte molecule with the metallic surface[16–19].

Tremendous progress in the area of Raman-based sensors led to the creation of surfaces allowing Raman signal detection from a few or even single molecules[20]. The SERS enhancement factor (EF), which is usually defined as

$$EF = \frac{I_{SERS}}{C_{SERS}} \frac{C_R}{I_R}, \qquad (1)$$

has been demonstrated to be as large as $10^5$-$10^{11}$ (cf. Ref. 5). Here, $I_{SERS}$ stands for the Raman intensity obtained for the SERS substrate under a certain analyte concentration $C_{SERS}$, and $I_R$ corresponds to the Raman intensity obtained under non-SERS conditions at analyte concentration $C_R$.

However, the adoption of SERS remains limited, to a great extent due to difficulties in fabrication of effective and affordable nanostructured plasmonic SERS surfaces. So far, SERS substrate fabrication techniques include lithography[7], laser ablation[21], templated electrodeposition[22], wet chemistry[23,24] and self-assembly[8,25]. The self-assembly technique is often considered the most promising because of its simplicity, cost- and time-effectiveness. For example, based on this approach, highly-branched dendritic nanostructures with sub-10-nm gaps between neighboring branches can be fabricated[26–28]. However, self-assembly techniques suffer from the lack of control over resulting structures. Thus, despite relatively satisfactory results[29], regularly reproducible detection of ultra-small concentrations of the analyte (in particular single molecules) using dendrites was not demonstrated. The main problem is related to the formation of bulk agglomerates[26,30–32], which do not allow stable detection at ultra-low concentrations of molecules due to the penetration of the analyte into the interior of such structures. As a result, the excitation laser does not access the analyte localization region and, consequently, the signal from the analyte is poorly recorded.

For that reason, such problem can be solved by the spatial separation of nanostructures using porous templates: polymer track membranes, anodized aluminum oxide, porous silicon, etc.[33–36]. The formation of metallic nanostructures in the pores matrix is usually carried out using two methods – electrochemical deposition and electroless galvanic displacement. The first method is used to produce nanostructures filling the pore volume (for example, in the formation of nanorods)[37], and the second – when it is required to realize individual nanoparticles or thin metallic layers on the pores surface[38].

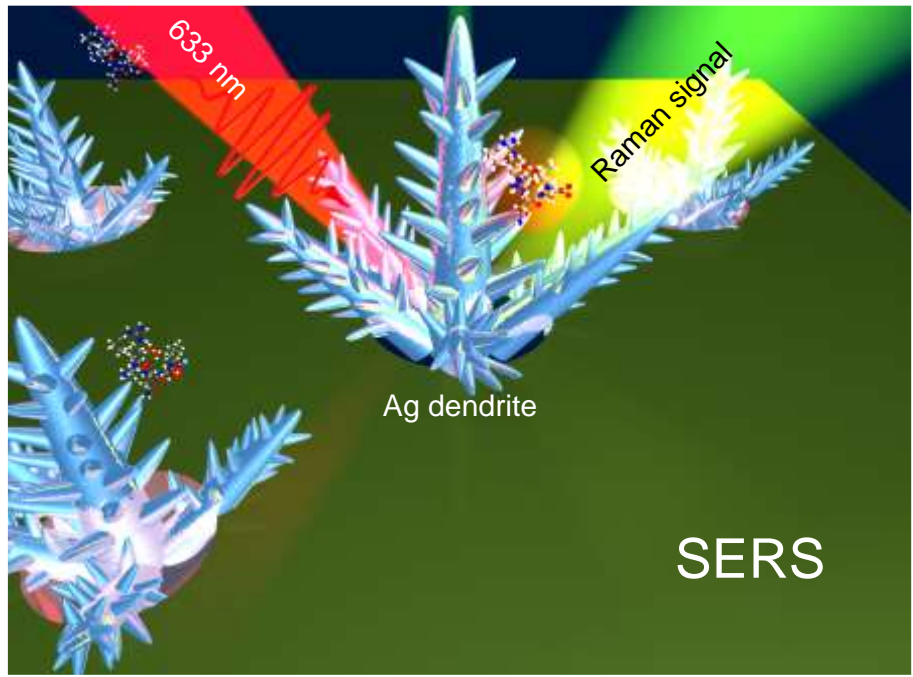

**Figure 1**. Schematic representation of enhanced Raman scattering from a molecule nearby a silver dendrite.

In this paper, we develop and demonstrate a novel technique for the fabrication of spatially separate silver dendrites, presented schematically in *Figure 1*, via self-organization of Ag nanostructures in pores of $SiO_2$/Si templates. We investigate the microstructure, morphology, optical features and SERS properties of fabricated nanostructures. Also, we show that observed templates allow us to achieve a SERS enhancement factor of ~$10^8$ with a single molecule detection possibility (~$10^{-15}$ M).

## Results and discussion

The formation of Ag nanostructures in the pores of $SiO_2$ template on the Si substrate was carried out by electroless galvanic displacement method using an aqueous solution of Ag nitrate ($AgNO_3$) and hydrofluoric acid (HF), which dissociates in water into individual cations and anions ($Ag^+$, $NO_3^-$, $H^+$, $F^-$) as was published in our previous publication[39]. The resulted ions participate in three parallel processes[40,41]: electrochemical reduction of $[Ag]^+$ ions to a metallic silver state on Si or metal [Eq. (2)] with simultaneous anodic and cathodic oxidation of Si [Eq. (3)], and also $SiO_2$ dissolution in hydrofluoric acid [Eq. (4)]. Schematically, the processes taking place in the pore of the $SiO_2$/Si template are shown *in Figure 2*.

$$4Ag^+ + Si + 6F^- \to 4Ag + SiF_6^{2-}, \quad Si + 6HF \to H_2SiF_6 + 4H^+ + 4e^-, \qquad (2)$$
$$Ag^+ + e^- \to Ag, \quad H_2 \to 2H^+ + 2e^-,$$

$$Si + 2H_2O \rightarrow SiO_2 + 4H^+ + 4e^-, \quad Si + 2HNO_3 \rightarrow NO + NO_2 + H_2O, \tag{3}$$

$$SiO_2 + 6HF \rightarrow H_2SiF_6 + 2H_2O. \tag{4}$$

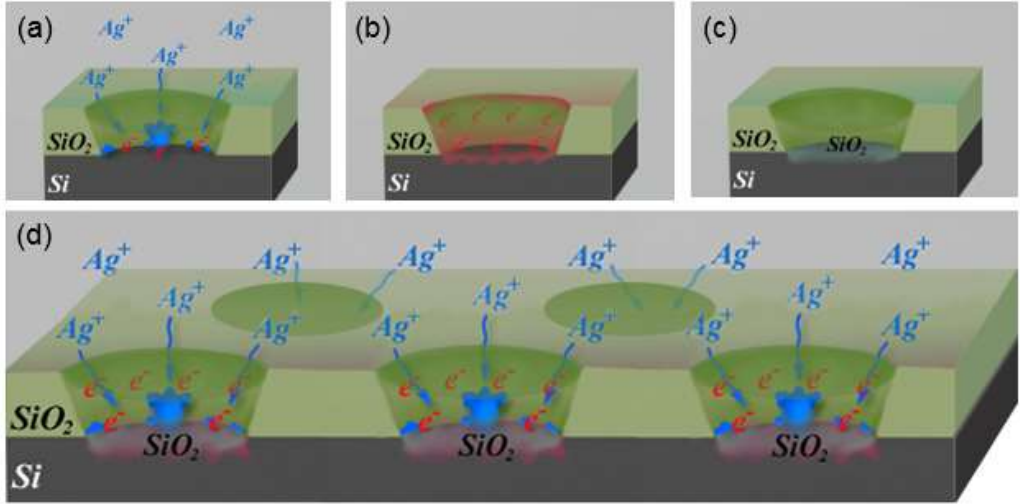

**Figure 2**. Schematic representation of the processes involved in reduction of $[Ag]^+$ ions occurring in pores of a template $SiO_2$ on Si: (a) Ag reduction; (b) $SiO_2$ etching; (c) Si oxidization; (d) all processes simultaneously.

All the pores of the ion-track $SiO_2$/p-Si template are filled with dendritic silver nanostructures as shown in *Figure 3a*. The SEM image of a single pore filled with silver nanostructures is presented in *Figure 3b*. The image reveals that the nanostructures have the form of dendrites containing the main trunk, lateral branches and many rounded outgrowths on them. Due to the presence of nanoscale inhomogeneities between dendrites branches, the formation of "hot spots" can be expected with the enhanced electric field, which can be used for boosting the SERS effect.

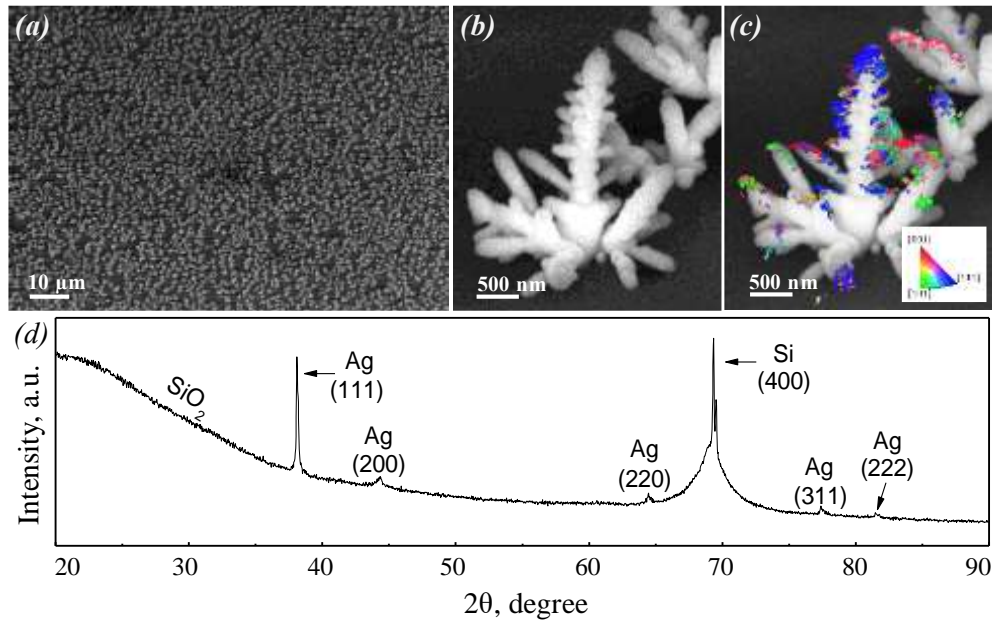

**Figure 3**. Morphology and microstructure of observed SiO$_2$(Ag)/ Si structures: (a, b) overview and single feature SEM micrographs, (c) EBSD analysis of silver dendrite, (d) X-ray diffraction pattern; lattice parameter (*a*) of Ag dendrites is 4.0816 Å.

The detailed structure of dendrites local features was evaluated by electron backscattered diffraction (EBSD). *Figure 3c* shows the EBSD orientation maps with inverse pole figure related color coding related to the sample surface normal direction z (IPFZ). For cubic systems, red color means that <100> crystal directions are parallel to the normal direction of the sample, if green or blue than <110> or <111> are aligned to normal direction, respectively. Our results suggest the presence of many blue spots around the main trunk that is corresponding to the <111> crystal growth direction. Further investigations of grain boundary relations inside the nanostructure reveal that most of the grain boundaries are coincidence site lattice (CSL[42]) grain boundaries Σ3 (twin grain boundary) and few Σ9 (for more details see Methods).

The crystallinity of the silver dendrites in the pores of the SiO$_2$/p-Si template was determined using X-ray diffraction method. *Figure 3d* shows the XRD pattern of the produced structure. Unlike to the typical diffraction patterns of silver dendrite[29,43], we observe a broad background signal coming from the amorphous SiO$_2$ layer. The peak at 2Theta about 69 deg corresponds to the signal from monocrystalline Si with highly broadened value due to the high defectiveness of the near-surface layers, which is caused by the consequence of creating a template using ion-track technology in the stage of irradiation with swift heavy ions. Ag dendrites have a cubic crystalline structure and are characterized by a set of reflexes from planes (111), (200), (220), (311) and (222). The broadened peak of (200) reflex, also, points to the polycrystalline nature of the observed Ag precipitate. The high intensity of (111) reflex indicates the priority direction of crystallite growth along this direction that correlates with EBSD observations in *Figure 3c*.

To investigate the optical properties of the fabricated dendritic Ag nanostructures numerically, we perform the full-wave simulations using CST Microwave Studio (*Figure 4a, b*). The simulation results indicate that the most intense hot spots are located between the neighboring dendrite branches, where the electric field enhancement reaches 12 as shown in *Figure 4b*. It is known that the SERS enhancement factor caused by electric field localization is roughly proportional to the fourth power of the electric field amplitude ($\sim E^4$). This gives us an estimation for the enhancement factor of $\sim 10^4$.

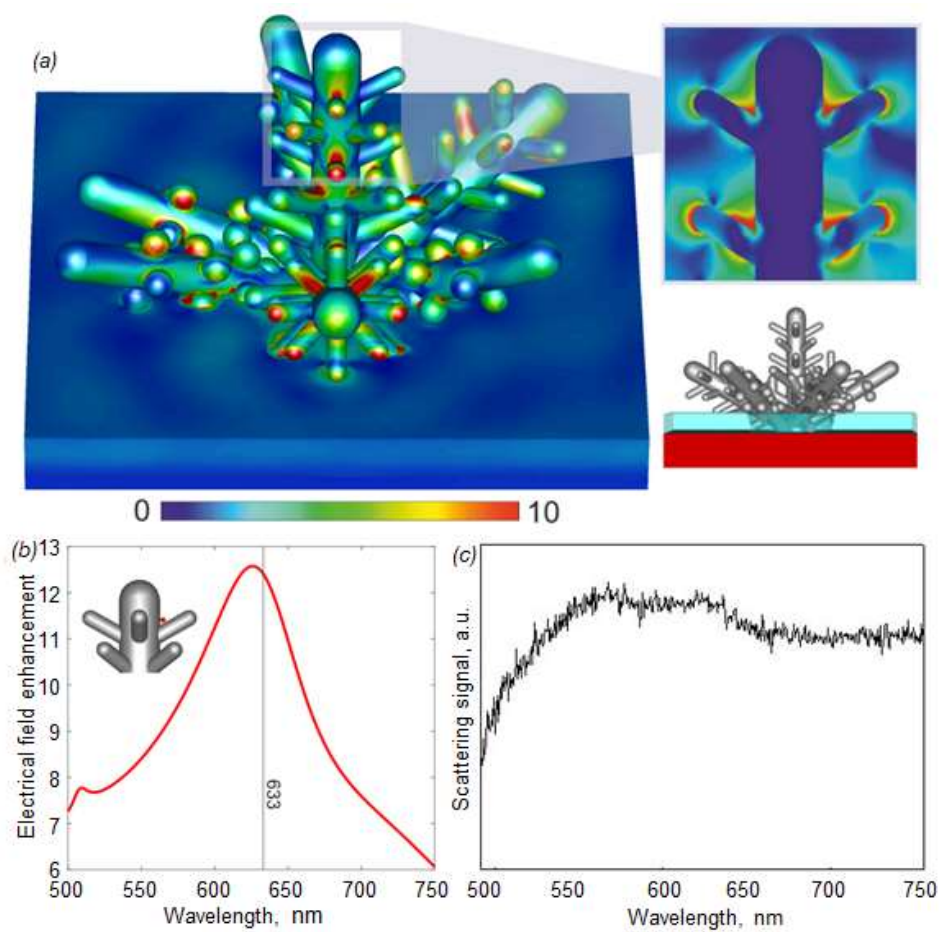

**Figure 4**. (a) Distribution of the electric field (logarithmic scale) over the surface of silver dendrite located in the pore of the $SiO_2$/ Si template. (b) Electrical field enhancement spectrum in the point indicated by inset. (c) Scattering spectrum of the dendritic nanostructures in the pores of the $SiO_2$/Si template.

The typical scattering spectrum (for more details see Methods) of the dendritic nanostructures is shown in *Figure 4c*. The complexness of the structure gives rise to spectral broadening of the measured scattering spectra[39]. Moreover, the scattered light collecting area has a radius of 15 μm, and hence each spectrum is averaged over ~50 nanostructures, which causes its additional broadening. In result, the scattering spectrum in *Figure 4c* does not exhibit sharp peaks in the range of 500 - 750 nm with a broadband resonant behavior nearby excitation wavelength (633 nm).

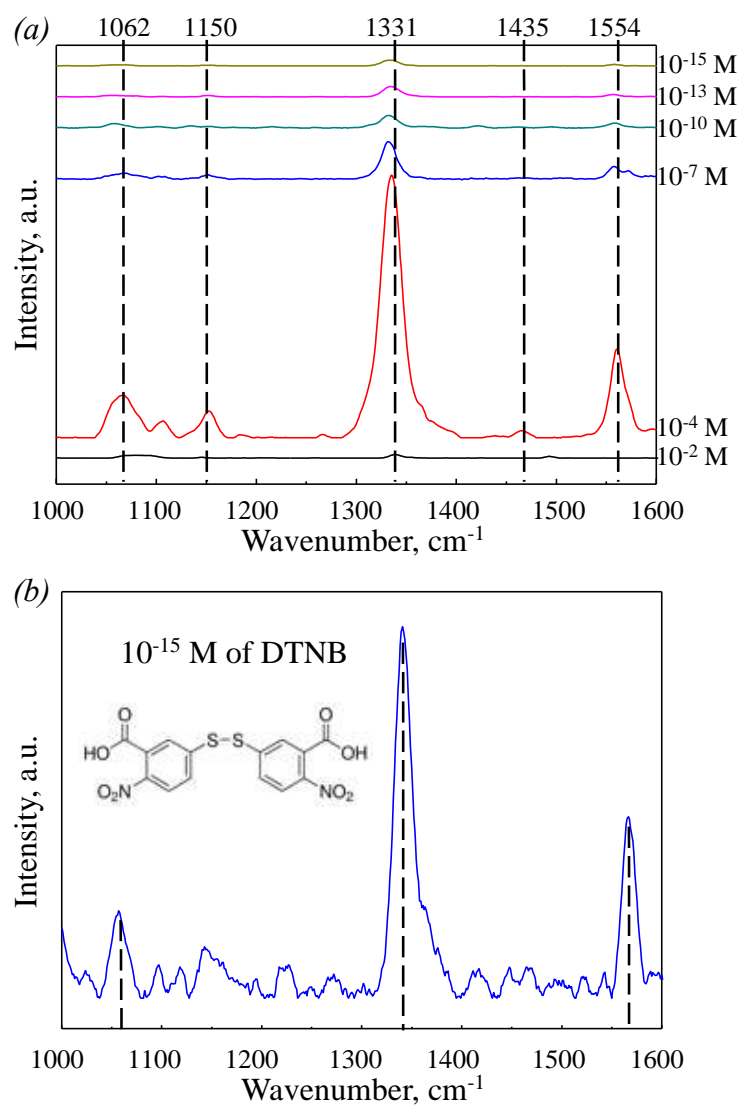

**Figure 5**. (a) Raman spectrum for a dried alcohol solution of DTNB with the concentration of $10^{-2}$ M (black curve), as well as SERS spectra obtained at $\lambda = 633$ nm excitation from spatially separated Ag dendrites for analyte concentrations of $10^{-4}$ M, $10^{-7}$ M, $10^{-10}$ M, $10^{-13}$ M, and $10^{-15}$ M. (b) Zoomed SERS spectrum of DNTB with the concentration of $10^{-15}$ M.

Next, in order to demonstrate the performance of the resulting structures as a sensor, the SERS measurements with using DTNB as a model analyte were performed (for details see Methods). *Figure 5* summarizes the obtained results. We attribute the observed most intense bands at 1062, 1150, 1331, 1554 cm$^{-1}$ to the DTNB[44] analyte. Evolution of SERS spectra shows the possibility of detection of ultra-low concentrations of the model DTNB analyte by applying localized silver dendrite nanostructures which correspond to EF at least $10^8$.

The EF value of $10^8$ exceeds by several orders of magnitude the results of numerical simulations. We believe that this indicates the existence of a strong chemical contribution to SERS enhancement. This contribution can be caused by the following mechanisms: a change in the polarization of the molecule[16,17] (static charge transfer, causing an amplification up to 10 times) and the transition of a charge from a molecule to a metal (or vice versa) with the formation of a new metal-molecule bond, analyte creates new allowed energy levels[17,19,45]. The energy difference of these levels coincides or lies near the energy of exciting wave, which leads to resonant Raman scattering. It is known that such a mechanism can provide the amplification of ~$10^2$-$10^3$ [46] and even more[47]. Thus, the cumulative gain factor $10^8$ is determined by the electromagnetic ($EF_{e.m.}$) and chemical ($EF_{chem}$) contributions[48,49]:

$$EF = EF_{e.m.} \times EF_{chem}. \qquad (5)$$

The appearance of new allowed levels in the energy spectrum of the molecule can lead to differences in comparison with the Raman spectrum of a bare analyte. Indeed, our results show (*Figure 5a*) that the main lines are shifted to the blue side and with decrease in the analyte concentration this effect gets enhanced. This behavior of the SERS spectra can be explained in the following way. With a sufficient amount of analyte molecules, both amplification mechanisms are involved: electromagnetic, which does not affect the spectrum change, and chemical, which causes the shift of the combination bands and is noticeably weaker than the electromagnetic one. With a decrease in the analyte concentration, the role of the electromagnetic mechanism is weakened, because fewer molecules get into "hot spots". Molecules, which still fall into these regions, fit closely to the metal and at distances of the order of several angstroms create a molecular-metal complex with silver. Thus, the molecule gets additional allowed energy levels. For that reason, the resonant transitions between these levels can also be enhanced by the electromagnetic mechanism. A decrease in concentration leads to the fact that molecules that fall into "hot spots" at distances of several tens of nanometers from a metal don't form a bond with silver surface. This is a consequence of the dominant effect of the chemical contribution to the SERS spectrum and a more noticeable shift in the combination lines. Note that the formation of new molecule-metal complexes can occur not only in "hot spots", but also in other regions of the metal, which also affects the characteristic shape of the SERS spectrum with a decrease of the electromagnetic mechanism role.

## Conclusions

Spatially separated Ag dendritic nanostructures in the pores of the ion-track template $SiO_2/Si$ were synthesized using directed self-organization of Ag in a limited volume. Microstructure and optical features of $SiO_2(Ag)/Si$ nanostructures have been studied numerically and experimentally. It has been shown that the Ag nanostructures have the form of dendrites. The SERS measurement results indicate the possibility of ultra-low analyte concentration detection, comparable to the concentrations of single molecules. The high sensitivity of the fabricated structures is justified by the coexistence of electromagnetic and chemical contributions to the Raman signal amplification. The modeling of the electromagnetic wave interaction with dendrites indicates that the appearance of "hot spots" occurs in places of twinning (intergrowth of the branches of the dendrite). The fabricated dendrites have a large number of "hot spots" with an electric field enhancement factor of 12. The random distribution of "hot spots" on a dendrite leads to its broadened scattering spectrum. We strongly believe that the simplicity, reproducibility, and scalability of the presented method for the formation of spatially separated dendrites make it promising for various vital applications in photonics, chemistry, and biology.

## Materials and methods

To obtain porous $SiO_2$ templates on Si substrate the ion-track technology was used [50,51]. For the substrate irradiation, single-crystal silicon (*p*-type, Boron doped, 12 Ohm·cm (100)) wafers were used with a layer of an amorphous $SiO_2$ dielectric layer with a thickness of about 700 nm. The wafer irradiation was performed by $^{152}Sm$ ions with the energy of 350 MeV and fluence of $10^8$ cm$^{-2}$ at UNILAC linear accelerator (GSI, Darmstadt, Germany). The low fluence was determined by the need to obtain spatially separated silver nanostructures. In details, the process of obtaining $SiO_2/Si$ templates is described in our previous work[35]. In our experiments, the etching was carried out in hydrofluoric acid (HF) at a concentration of 1%. The total irradiated surface treatment time was 110 min, which provided a pore diameter of ~ 800 nm.

The growth of silver nanostructures in porous $SiO_2/Si$ templates was carried out using the electroless galvanic displacement method. Firstly, the templates were washed in isopropyl alcohol, and then in distilled water and dried in air for 30 minutes. The electrolyte for the silver deposition was an aqueous 0.02M solution of silver nitrate ($AgNO_3$) and 5M hydrofluoric acid (HF). The deposition of metal into the pores was carried out at a temperature of 35 °C for 30 s. Finally, the samples were washed in distilled water and dried in a stream of nitrogen.

The scattering properties (LSPR - Localized Surface Plasmon Resonance) of the metal nanostructures were recorded with a dark-field micro-spectroscopy similarly as in Ref.[52]. Briefly, the micro-spectroscopy system consisting of an upright microscope (AxioImager Z1m, Carl Zeiss) combined with a fiber-coupled spectrometer (Spectra Pro 2300i, Princeton Instruments) and used in the dark-field top illumination settings. The microscope was equipped with a color camera (AxioCam Mrc5, Carl Zeiss) to record the overview images of the samples. The scattering signal was recorded from a circular area with the diameter of 15 μm (10x objective; NA = 0.2, 150 μm diameter aperture in front of the collecting fiber) from the middle of the overview image. The scattering signals were corrected for a background and a lamp spectrum.

To clarify the crystal structure of silver dendrites, the X-Ray diffraction (XRD) patterns were measured using X'Pert Pro X-ray diffractometer from PANalytical B.V. measured in the Bragg-Brentano arrangement.

Scanning electron microscopy (SEM, Carl Zeiss ULTRA 55, FE-SEM) was chosen to characterize the surface morphology of the obtained samples.

The EBSD (Electron Back-Scatter Diffraction) analysis of the samples was performed using a scanning electron microscope equipped with a Schottky emitter (Zeiss Ultra 55) and a high resolution EBSD camera (Bruker e-Flash[HR]) using an acceleration voltage of 20 kV at a nominal sample current of 1.5 nA, sample tilt of 60° with respect to the primary electron beam, and a scanning step width of 12 nm. The resolution of the Kikuchi patterns was 160×120 pixels applying a binning of 10×10 and using integration times of 40 ms. There were no problems with charging issues. The data were evaluated using a commercial software package (Bruker Quantax; Esprit 2.1) as well as the MATLAB toolbox MTEX[53].

The SERS measurements were performed using commercially available confocal micro-spectrometer "Confotec CARS" (SOL Instruments Ltd., Belarus) setup equipped with several laser systems including a 633 nm laser[54]. During the measurements a 600 lines per mm grating was used with a spectral resolution of ~ 2 cm$^{-1}$, the same objective (Leica 100x, 0.9 N.A.) was employed for focusing the laser beam on the sample and for collecting the backscattered light. The laser power 50 μW within the exposure time of 1s was exploited. For the SERS studies, the Ellman's reagent $C_{14}H_8N_2O_8S_2$ (5,50-dithiobis-(2-nitrobenzoic acid)) or DTNB was used as a model analyte in different concentrations: $10^{-4}$ M, $10^{-7}$ M, $10^{-10}$ M, $10^{-13}$ M, and $10^{-15}$ M. For registration of ultra-low concentrations ($10^{-13}$ M and $10^{-15}$ M) of the analyte, 10×10 μm mapping was used. For comparing with SERS spectra Raman spectrum measured from $10^{-2}$ M of DTNB on the $SiO_2$/Si template without silver.

CST Microwave Studio is a full-wave 3D electromagnetic field solver based on finite-integral time domain (FITD) solution technique. A nonuniform mesh was used to improve accuracy near the Ag nanorods where the field concentration was significantly large[55]. The $SiO_2$ substrate thickness has been taken to be 170 nm. The lengths of big and small dendritic were ~1130 nm and 170 nm, respectively. The thickness of the big and small dendrites was 180 nm and 70 nm, respectively. The radius of the rounded outgrowths was 53 nm.

# References

(1)   Krafft, C.; Schie, I. W.; Meyer, T.; Schmitt, M.; Popp, J. Developments in Spontaneous and Coherent Raman Scattering Microscopic Imaging for Biomedical Applications. *Chem. Soc. Rev.* **2016**, *45* (7), 1819–1849.

(2)   Moskovits, M. Surface-Enhanced Raman Spectroscopy: A Brief Retrospective. *Journal of Raman Spectroscopy* **2005**, *36*, 485–496.

(3)   Kneipp, K.; Kneipp, H.; Itzkan, I.; Dasari, R. R.; Feld, M. S. Surface-Enhanced Raman Scattering and Biophysics. *Journal of Physics: Condensed Matter* **2002**, *14*, R597–R624.

(4)   Höppener, C.; Novotny, L. Exploiting the Light–Metal Interaction for Biomolecular Sensing and Imaging. *Quarterly Reviews of Biophysics* **2012**, *45* (02), 209–255.

(5)   Balčytis, A.; Nishijima, Y.; Krishnamoorthy, S.; Kuchmizhak, A.; Stoddart, P. R.; Petruškevičius, R.; Juodkazis, S. From Fundamental toward Applied SERS: Shared Principles and Divergent Approaches. *Advanced Optical Materials* **2018**, 1800292.

(6)   Ringe, E.; Sharma, B.; Henry, A.-I.; Marks, L. D.; Van Duyne, R. P. Single Nanoparticle Plasmonics. *Physical Chemistry Chemical Physics* **2013**, *15* (12), 4110.

(7)   Gopalakrishnan, A.; Chirumamilla, M.; De Angelis, F.; Toma, A.; Zaccaria, R. P.; Krahne, R. Bimetallic 3D Nanostar Dimers in Ring Cavities: Recyclable and Robust Surface-Enhanced Raman Scattering Substrates for Signal Detection from Few Molecules. *ACS Nano* **2014**, *8* (8), 7986–7994.

(8)   Gwo, S.; Wang, C. Y.; Chen, H. Y.; Lin, M. H.; Sun, L.; Li, X.; Chen, W. L.; Chang, Y. M.; Ahn, H. Plasmonic Metasurfaces for Nonlinear Optics and Quantitative SERS. *ACS Photonics* **2016**, *3* (8), 1371–1384.

(9)   Santoro, G.; Yu, S.; Schwartzkopf, M.; Zhang, P.; Koyiloth Vayalil, S.; Risch, J. F. H.; Rübhausen, M. A.; Hernández, M.; Domingo, C.; Roth, S. V. Silver Substrates for Surface Enhanced Raman Scattering: Correlation between Nanostructure and Raman Scattering Enhancement. *Applied Physics Letters* **2014**, *104* (24), 243107.

(10) Kneipp, K. Surface-Enhanced Raman Scattering. *Physics Today* **2007**, *60* (11), 40–46.

(11) Xia, Y.; Campbell, D. J. Plasmons: Why Should We Care? *Journal of Chemical Education* **2007**, *84* (1), 91.

(12) Kaniukov, E.; Yakimchuk, D.; Arzumanyan, G.; Terryn, H.; Baert, K.; Kozlovskiy, A.; Zdorovets, M.; Belonogov, E.; Demyanov, S. Growth Mechanisms of Spatially Separated Copper Dendrites in Pores of a SiO 2 Template. *Philosophical Magazine* **2017**, *6435*, 1–16.

(13) Shahbazyan, T. V; Stockman, M. I. *Plasmonics: Theory and Applications*; Shahbazyan, T. V., Stockman, M. I., Eds.; Springer Netherlands: Dordrecht, 2013.

(14) Sharma, B.; Frontiera, R. R.; Henry, A.-I.; Ringe, E.; Van Duyne, R. P. SERS: Materials, Applications, and the Future. *Materials Today* **2012**, *15* (1–2), 16–25.

(15) Krasnok, A.; Caldarola, M.; Bonod, N.; Alú, A. Spectroscopy and Biosensing with Optically Resonant Dielectric Nanostructures. *Advanced Optical Materials* **2018**, *6* (5), 1701094.

(16) Kneipp, K. Chemical Contribution to SERS Enhancement: An Experimental Study on a Series of Polymethine Dyes on Silver Nanoaggregates. *Journal of Physical Chemistry C* **2016**, *120* (37), 21076–21081.

(17) Zhao, L. L.; Jensen, L.; Schatz, G. C. Pyridine-Ag20 Cluster: A Model System for Studying Surface-Enhanced Raman Scattering. *J. Am. Chem. Soc.* **2006**, *128* (9), 2911–2919.

(18) Arenas, J. F.; Woolley, M. S.; Tocón, I. L.; Otero, J. C.; Marcos, J. I. Complete Analysis of the Surface-Enhanced Raman Scattering of Pyrazine on the Silver Electrode on the Basis of a Resonant Charge Transfer Mechanism Involving Three States. *The Journal of Chemical Physics* **2000**, *112* (17), 7669–7683.

(19) Xia, L.; Chen, M.; Zhao, X.; Zhang, Z.; Xia, J.; Xu, H.; Sun, M. Visualized Method of Chemical Enhancement Mechanism on SERS and TERS. *Journal of Raman Spectroscopy* **2014**, *45* (7), 533–540.

(20) Nie, S. Probing Single Molecules and Single Nanoparticles by Surface-Enhanced Raman Scattering. *Science* **1997**, *275* (1997), 1102–1106.

(21) Makarov, S. V.; Milichko, V. A.; Mukhin, I. S.; Shishkin, I. I.; Zuev, D. A.; Mozharov, A.

M.; Krasnok, A. E.; Belov, P. A. Controllable Femtosecond Laser-Induced Dewetting for Plasmonic Applications. *Laser & Photonics Reviews* **2016**, *10* (1), 91–99.

(22) Abdelsalam, M. E.; Mahajan, S.; Bartlett, P. N.; Baumberg, J. J.; Rusell, A. E. SERS at Structured Palladium and Platinum Surfaces. *Journal of the American Chemical Society* **2007**, *129* (23), 7399–7406.

(23) Panarin, A. Y.; Terekhov, S. N.; Kholostov, K. I.; Bondarenko, V. P. SERS-Active Substrates Based on n-Type Porous Silicon. *Applied Surface Science* **2010**, *256* (23), 6969–6976.

(24) Bandarenka, H.; Artsemyeva, K.; Redko, S.; Panarin, A.; Terekhov, S.; Bondarenko, V. Effect of Swirl-like Resistivity Striations in N+-Type Sb Doped Si Wafers on the Properties of Ag/Porous Silicon SERS Substrates. *Physica Status Solidi (C)* **2013**, *10* (4), 624–627.

(25) Dong, D.; Yap, L. W.; Smilgies, D. M.; Si, K. J.; Shi, Q.; Cheng, W. Two-Dimensional Gold Trisoctahedron Nanoparticle Superlattice Sheets: Self-Assembly, Characterization and Immunosensing Applications. *Nanoscale* **2018**.

(26) Brejna, P. R.; Griffiths, P. R. Electroless Deposition of Silver Onto Silicon as a Method of Preparation of Reproducible Surface-Enhanced Raman Spectroscopy Substrates and Tip-Enhanced Raman Spectroscopy Tips. *Applied Spectroscopy* **2010**, *64* (5), 493–499.

(27) Qiu, T.; Wu, X. L.; Mei, Y. F.; Chu, P. K.; Siu, G. G. Self-Organized Synthesis of Silver Dendritic Nanostructures via an Electroless Metal Deposition Method. *Applied Physics A: Materials Science and Processing* **2005**, *81* (4), 669–671.

(28) Qiu, T.; Wu, X. L.; Shen, J. C.; Xia, Y.; Shen, P. N.; Chu, P. K. Silver Fractal Networks for Surface-Enhanced Raman Scattering Substrates. *Applied Surface Science* **2008**, *254* (17), 5399–5402.

(29) Alam, M. M.; Ji, W.; Luitel, H. N.; Ozaki, Y.; Watari, T.; Nakashima, K. Template Free Synthesis of Dendritic Silver Nanostructures and Their Application in Surface-Enhanced Raman Scattering. *RSC Adv.* **2014**, *4* (95), 52686–52689.

(30) Ye, W.; Shen, C.; Tian, J.; Wang, C.; Hui, C.; Gao, H. Controllable Growth of Silver Nanostructures by a Simple Replacement Reaction and Their SERS Studies. *Solid State Sciences* **2009**, *11* (6), 1088–1093.

(31) Sun, X.; Lin, L.; Li, Z.; Zhang, Z.; Feng, J. Novel Ag–Cu Substrates for Surface-Enhanced Raman Scattering. *Materials Letters* **2009**, *63* (27), 2306–2308.

(32) Senthil Kumaran, C. K.; Agilan, S.; Dhayalan Velauthapillai; Muthukumarasamy, N.; Thambidurai, M.; Ranjitha, A.; Balasundaraprabhu, R.; Senthil, T. S. Preparation and Characterization of Copper Dendrite Like Structure by Chemical Method. *Advanced Materials Research* **2013**, *678* (MARCH), 27–31.

(33) Hurst, S. J.; Payne, E. K.; Qin, L.; Mirkin, C. A. Multisegmented One-Dimensional Nanorods Prepared by Hard-Template Synthetic Methods. *Angewandte Chemie International Edition* **2006**, *45* (17), 2672–2692.

(34) Masuda, H.; Fukuda, K. Ordered Metal Nanohole Arrays Made by a Two-Step Replication of Honeycomb Structures of Anodic Alumina. *Science* **1995**, *268* (5216), 1466–1468.

(35) Kaniukov, E. Y.; Ustarroz, J.; Yakimchuk, D. V; Petrova, M.; Terryn, H.; Sivakov, V.; Petrov, A. V. Tunable Nanoporous Silicon Oxide Templates by Swift Heavy Ion Tracks Technology. *Nanotechnology* **2016**, *27* (11), 115305.

(36) Bandarenka, H.; Girel, K.; Zavatski, S.; Panarin, A.; Terekhov, S. Progress in the Development of SERS-Active Substrates Based on Metal-Coated Porous Silicon. *Materials* **2018**, *11* (5), 852.

(37) Cao, Y.; Mallouk, T. E. Morphology of Template-Grown Polyaniline Nanowires and Its Effect on the Electrochemical Capacitance of Nanowire Arrays. *Chemistry of Materials* **2008**, *20* (16), 5260–5265.

(38) Wirtz, M.; Martin, C. R. Template-Fabricated Gold Nanowires and Nanotubes. *Advanced Materials* **2003**, *15* (5), 455–458.

(39) Yakimchuk, D.; Kaniukov, E.; Bundyukova, V.; Osminkina, L.; Teichert, S.; Demyanov, S.; Sivakov, V. Silver Nanostructures Evolution in Porous SiO2/p-Si Matrices for Wide Wavelength Surface-Enhanced Raman Scattering Applications. *MRS Communications* **2018**, *8* (1), 95–99.

(40) Abouda-Lachiheb, M.; Nafie, N.; Bouaicha, M. The Dual Role of Silver during Silicon Etching in HF Solution. *Nanoscale research letters* **2012**, *7* (1), 455.

(41) Ye, W.; Chang, Y.; Ma, C.; Jia, B.; Cao, G.; Wang, C. Electrochemical Investigation of the

Surface Energy: Effect of the HF Concentration on Electroless Silver Deposition onto p-Si (111). *Applied Surface Science* **2007**, *253* (7), 3419–3424.

(42) Sutton;, A. P.; Balluffi, R. W. *Interfaces in Crystalline Materials*; Oxford : Clarendon Press: New York, 1995.

(43) He, R.; Qian, X.; Jie, Y.; Zhu, Z. Formation of Silver Dendrites under Microwave Irradiation. *Chemical Physsics Letters* **2003**, *369*, 454–458.

(44) Lin, C. C.; Chang, C. W. AuNPs@mesoSiO2 Composites for SERS Detection of DTNB Molecule. *Biosensors and Bioelectronics* **2014**, *51*, 297–303.

(45) Arenas, J. F.; Woolley, M. S.; Tocon, I. L.; Otero, J. C.; Marcos, J. I. Complete Analysis of the Surface-Enhanced Raman Scattering of Pyrazine on the Silver Electrode on the Basis of a Resonant Charge Transfer Mechanism Involving Three States. *Journal Of Chemical Physics* **2000**, *112* (17), 7669–7683.

(46) Saikin, S. K.; Olivares-Amaya, R.; Rappoport, D.; Stopa, M.; Aspuru-Guzik, A. On the Chemical Bonding Effects in the Raman Response: Benzenethiol Adsorbed on Silver Clusters. *Physical Chemistry Chemical Physics* **2009**, *11*, 9401–9411.

(47) Fromm, D. P.; Sundaramurthy, A.; Kinkhabwala, A.; Schuck, P. J.; Kino, G. S.; Moerner, W. E. Exploring the Chemical Enhancement for Surface-Enhanced Raman Scattering with Au Bowtie Nanoantennas. *Journal of Chemical Physics* **2006**, *124* (6).

(48) Saikin, S. K.; Chu, Y.; Rappoport, D.; Crozier, K. B.; Aspuru-Guzik, A. Separation of Electromagnetic and Chemical Contributions to Surface-Enhanced Raman Spectra on Nanoengineered Plasmonic Substrates. *Journal of Physical Chemistry Letters* **2010**, *1* (18), 2740–2746.

(49) Otto, A. The "chemical" (Electronic) Contribution to Surface-Enhanced Raman Scattering. *Journal of Raman Spectroscopy* **2005**, *36* (6–7), 497–509.

(50) Fink, D.; Petrov, A. V.; Hoppe, K.; Fahrner, W. R.; Papaleo, R. M.; Berdinsky, A. S.; Chandra, A.; Chemseddine, A.; Zrineh, A.; Biswas, A.; et al. Etched Ion Tracks in Silicon Oxide and Silicon Oxynitride as Charge Injection or Extraction Channels for Novel Electronic Structures. *Nuclear Instruments and Methods in Physics Research, Section B: Beam Interactions with Materials and Atoms* **2004**, *218* (1–4), 355–361.

(51) Dallanora, A.; Marcondes, T. L.; Bermudez, G. G.; Fichtner, P. F. P.; Trautmann, C.; Toulemonde, M.; Papaléo, R. M. Nanoporous SiO2/Si Thin Layers Produced by Ion Track Etching: Dependence on the Ion Energy and Criterion for Etchability. *Journal of Applied Physics* **2008**, *104* (2), 024307-1–024307-8.

(52) Jahr, N.; Anwar, M.; Stranik, O.; Hädrich, N.; Vogler, N.; Csaki, A.; Popp, J.; Fritzsche, W. Spectroscopy on Single Metallic Nanoparticles Using Subwavelength Apertures. *The Journal of Physical Chemistry C* **2013**, *117* (15), 7751–7756.

(53) Krakow, R.; Bennett, R. J.; Johnstone, D. N.; Vukmanovic, Z.; Solano-Alvarez, W.; Lainé, S. J.; Einsle, J. F.; Midgley, P. A.; Rae, C. M. F.; Hielscher, R. On Three-Dimensional Misorientation Spaces. *Proceedings of the Royal Society A: Mathematical, Physical and Engineering Science* **2017**, *473* (2206), 20170274.

(54) Fabelinsky, V. I.; Kozlov, D. N.; Orlov, S. N.; Polivanov, Y. N.; Shcherbakov, I. A.; Smirnov, V. V.; Vereschagin, K. A.; Arzumanyan, G. M.; Mamatkulov, K. Z.; Afanasiev, K. N.; et al. Surface-Enhanced Micro-CARS Mapping of a Nanostructured Cerium Dioxide/Aluminum Film Surface with Gold Nanoparticle-Bound Organic Molecules. *Journal of Raman Spectroscopy* **2018**, *49* (7), 1145–1154.

(55) Johnson, P. B.; Christy, R. W. Optical Constants of the Noble Metals. *Physical Review B* **1972**, *6* (12), 4370–4379.